\documentclass[runningheads]{llncs}
\usepackage{algorithm}
\usepackage[noend]{algpseudocode}
\usepackage{graphicx}
\usepackage{textcomp}
\usepackage{xcolor}
\usepackage{booktabs, multirow}
\usepackage{subfigmat}
\usepackage{tcolorbox}
\usepackage{color,soul}
\usepackage{paralist}
\usepackage{lipsum}

\makeatletter
\renewcommand{\fnum@algorithm}{\scriptsize{\textbf{\ALG@name~\thealgorithm}}}
\makeatother

\newcommand{\mypara}[1]{\vspace{2pt}\noindent{\bf{#1}}}

\author{Michael Wrana, Uzma Maroof, Diogo Barradas.}
\institute{University of Waterloo, Ontario, Canada\\
\email{\{mmwrana,uzma.maroof,diogo.barradas\}@uwaterloo.ca}}

\begin{document}

\title{TSA-WF: Exploring the Effectiveness of Time Series Analysis for Website Fingerprinting}

\titlerunning{The Effectiveness of Time Series Analysis for Website Fingerprinting}

\maketitle

\begin{abstract}

Website fingerprinting (WF) is a technique that allows an eavesdropper to determine the website a target user is accessing by inspecting the metadata associated with the packets she exchanges via some encrypted tunnel, e.g., Tor. Recent WF attacks built using machine learning (and deep learning) process and summarize trace metadata during their feature extraction phases.  This methodology leads to predictions that lack information about the instant at which a given website is detected within a (potentially large) network trace comprised of multiple sequential website accesses -- a setting known as \textit{multi-tab} WF.

In this paper, we explore whether classical time series analysis techniques can be effective in the WF setting.  Specifically, we introduce TSA-WF, a pipeline designed to closely preserve network traces' timing and direction characteristics, which enables the exploration of algorithms designed to measure time series similarity in the WF context. Our evaluation with Tor traces reveals that TSA-WF achieves a comparable accuracy to existing WF attacks in scenarios where website accesses can be easily singled-out from a given trace (i.e., the \textit{single-tab} WF setting), even when shielded by specially designed WF defenses. Finally, while TSA-WF did not outperform existing attacks in the multi-tab setting, we show how TSA-WF can help pinpoint the approximate instant at which a given website of interest is visited within a multi-tab trace.\footnote{This preprint has not undergone any post-submission improvements or corrections. The Version of Record of this contribution is published in the Proceedings of the 20th International Conference on Availability, Reliability and Security (ARES 2025)}

\end{abstract}

\pagebreak

\section{Introduction}
Privacy-conscious internet users often seek to hide their web activities from network surveillance apparatuses.  Technologies such as Tor~\cite{DBLP:conf/uss/DingledineMS04} and VPNs~\cite{DBLP:conf/imc/KhanDVSKV18} respond to users' privacy needs by providing anonymous and encrypted access to websites.  Instead of directly accessing a website, Tor and VPNs 
rely on intermediate nodes that bridge the communication between clients and their intended destinations. Thus, an adversary that simply inspects packets' source/destination IP addresses cannot identify the website a target user is accessing.  

Unfortunately, the above privacy-enhancing technologies do not sufficiently hide the network metadata (e.g., packet sizes and timing, or the overall volume of communication) associated with a user's connection~\cite{DBLP:journals/popets/VeichtRB23}. For this reason, existing research has shown that an eavesdropping adversary is capable of determining the websites that a target user is visiting by comparing her traffic patterns with the patterns generated when accessing a pool of websites of interest. This method, known as website fingerprinting (WF), poses significant privacy risks~\cite{DBLP:journals/corr/abs-1708-06376}.

WF techniques have existed for nearly two decades~\cite{DBLP:conf/sp/SunSWRPQ02} but struggled to achieve satisfactory performance for real world usage~\cite{DBLP:conf/ccs/JuarezAADG14}.  Recent advances in machine learning (ML) and deep learning (DL) enabled eavesdropping adversaries to create sophisticated models that can accurately identify websites while requiring less training data~\cite{DBLP:conf/codaspy/WangDL0W21}, or in the early stages of page loading~\cite{10.1145/3658644.3670272}.  As of today, WF attacks  can achieve high success rates against Tor~\cite{DBLP:conf/ccs/SirinamIJW18} or VPNs~\cite{DBLP:journals/corr/abs-2403-03998}, and can even detect accesses to multiple websites in sequence or simultaneously (i.e., a setting known as \textit{multi-tab} WF)~\cite{DBLP:conf/sp/DengYLZLXXW23,DBLP:conf/ccs/JinLLS23}.  

To prepare a WF attack, the adversary must first collect traces of the websites it wishes to monitor. These traces are typically represented as a time series comprised of a sequence of IP packets organized according to arrival time, and annotated with the packet's direction. Then, the adversary extracts a set of features that describe each trace (i.e., a fingerprint) to prepare the training of a  classification model. To this end, the adversary will either use latent features extracted directly from the trace (e.g., with DL) or resort to manual feature engineering (e.g., by determining a set of summary statistics for each trace, such as the number of incoming and outgoing packets). Finally, the adversary can then use the trained model to match the fingerprint of a target user's network trace to the fingerprints of the websites the adversary is monitoring.

While adversaries lose valuable information about traces when performing manual feature engineering (e.g., abstracting away per-packet timing information)~\cite{DBLP:journals/popets/VeichtRB23}, the dimensionality reduction carried out by latent feature extraction procedures has so far resulted in DL classifiers that lack some insights over their predictions. For instance, existing DL attacks can predict the websites which are likely to be contained within a multi-tab trace, but cannot isolate the per-tab traces they issue these predictions for, being unable to specify the exact instant when a website was accessed~\cite{DBLP:conf/sp/DengYLZLXXW23} (i.e., they show a lack of \textit{temporal resolution}).

Our work departs from the observation that, despite the existence of numerous refined techniques for the analysis of time series data~\cite{middlehurst2024bake}, the usefulness of such techniques  has only been briefly touched upon within the WF context~\cite{DBLP:conf/sp/RupprechtKHP19}. Interestingly, these ``classical'' time series analysis algorithms may pose themselves as a suitable alternative to compare network packet traces and avoid the aforementioned pitfalls associated with manual feature engineering or latent feature extraction performed by current WF attacks.

In this paper, we comprehensively explore the potential of classical time series analysis techniques (specifically, those relying on the \textit{similarity} between time series, such as euclidean distance and dynamic time warping~\cite{muller2007dynamic}) to act as the main driver for matching website traces in the context of WF attacks. To this end, we devise TSA-WF, a time series analysis pipeline tailored to WF comprised of three components: a) a distance calculator that combines different time series analysis techniques to quantitatively determine the similarity between website traces; b) a classifier which uses distance scores to execute a WF attack, and; c) a tool that provides additional context to WF attacks' results by pinpointing the approximate instant at which a monitored website is accessed within a multi-tab trace, thus compensating for the information loss incurred by current attacks during their feature extraction phase.

The evaluation we conducted using recent Tor traces indicates that TSA-WF, when used as an independent attack, is on par with state-of-the-art WF classifiers in the single-tab setting, achieving a classification accuracy of 91.2\% for undefended traces. While TSA-WF is subpar compared to existing DL methods when applied to merged sequences of traces (i.e., the multi-tab setting), TSA-WF can be combined with existing attacks to provide valuable insights on why websites are (mis-)classified in this setting. In particular, in the \texttt{3-Tab} setting, TSA-WF can pinpoint the approximate location of a monitored website within the multi-tab traces to within 10k packets 83.7\% of the time.

\mypara{Contributions.} We summarize our main contributions as:

\begin{compactitem}
\item We characterize encrypted website traces as time series, and then frame WF attacks as a time series matching problem.
\item We design TSA-WF~\cite{anon}, a time series analysis pipeline geared towards WF which is compatible with multiple time series matching algorithms.
\item We evaluate TSA-WF resorting to multiple time series similarity metrics, and compare its accuracy with that of existing WF attacks in the single-tab and multi-tab settings using merged, overlapped, and defended Tor traces.
\item We show how TSA-WF augments the capabilities of WF attacks by pinpointing the approximate location of specific website traces in multi-tab settings.
\end{compactitem}

\section{Background} 

\subsection{The WF Threat Model}
\label{sec:threat-model}
 
\begin{figure}[t!]
    \centering
    \includegraphics[width=0.75\linewidth]{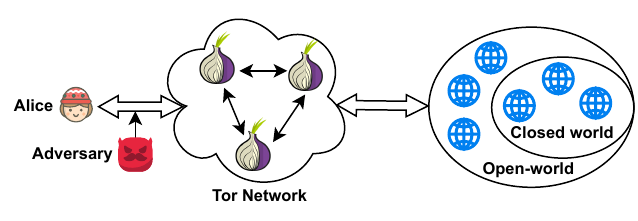}
    \vspace{-0.4cm}
    \caption{A standard website fingerprinting threat model over Tor. A local adversary eavesdrops Alice's encrypted communications while she accesses a set of websites.}
    \label{fig:threatmodel}
\end{figure}

The typical threat model for a WF attack (shown in Fig.~\ref{fig:threatmodel}) involves a user, Alice, attempting to privately browse the internet over Tor while an adversary aims to determine the sequence of websites she is visiting without cryptographically breaking her communication. To prepare a  WF attack, the adversary first builds a database of \textit{fingerprints} for the websites (e.g., Alexa top-100) it wishes to monitor before Alice visits them. Then, the adversary launches the attack by comparing its database of monitored website traces with the traces observed in the network once Alice has visited a given website. WF attacks are typically launched in two different scenarios and settings, which we discuss below.

\mypara{Closed- vs. open-world scenario.} In a \textit{closed-world} scenario, the assumption is that Alice visits a website which is amongst a limited set of websites monitored by the adversary, and for which the adversary collects sample traces as part of its database \cite{DBLP:journals/popets/VeichtRB23}.  In an \textit{open-world} scenario, Alice is allowed to access both monitored and unmonitored websites (i.e., the web at large).  Thus, the closed-world represents the best-case scenario for the adversary, while the open-world is considered to be more realistic. As it will become clearer in \S\ref{sec:experiment}, our evaluation focuses primarily on the open-world scenario.

\mypara{Single- vs. multi-tab setting.} The adversary operates in the \textit{single-tab} setting if it knows when Alice starts and stops loading each website.  In turn, we say that the adversary operates in the \textit{multi-tab} setting if it can only see a merged trace, potentially containing packets from multiple websites accessed in sequence or simultaneously overlapping~\cite{DBLP:conf/acsac/XuWLGCJ18}. The single-tab setting is advantageous for the adversary because a target user's traces can be directly compared with traces contained in the adversary's fingerprint database. Instead, in the multi-tab setting, the adversary must determine whether, when, and how many times Alice visited a monitored website in each of her merged traces. Our evaluation of TSA-WF (see \S\ref{sec:experiment}) addresses both the single- and multi-tab scenarios.

\subsection{Data Transformations in WF}
\label{sec:transformations}

To fuel higher WF attack accuracy, the raw network traces observed when accessing a website via some encrypted tunnel are converted into features suitable for training machine learning algorithms. Veicht \textit{et al.} \cite{DBLP:journals/popets/VeichtRB23} describe two transformations applied to raw traces before they are used to train a WF classifier.

\mypara{Raw packet representation.} Each raw packet in a website trace contains a wealth of information about an ongoing data transmission (e.g., packet arrival times, order, size, and TCP/IP header fields).  In the first transformation, the data from every packet is reduced into a trace representation, containing direction (i.e., incoming vs outgoing) and timing (e.g., how many milliseconds have elapsed between the start of recording and when each packet arrived). Note that packet size is usually ignored in the Tor context, because Tor exchanges data using same-sized cells \cite{DBLP:conf/wpes/WangG13}. This first transformation is fueled by the fact that many raw packet features are either redundant or cause overfitting \cite{DBLP:journals/popets/YanK18}.

\mypara{Trace representation.} This representation of a website access, first proposed by Wang \textit{et al.} \cite{DBLP:conf/wpes/WangG13} is essentially a time series, as the packets are strictly ordered by their arrival time.  The trace representation consists of a list of integers where each packet has a sign for direction and magnitude for arrival time (e.g., \texttt{[-2.4]} means an incoming packet recorded after 2.4ms).  In the second transformation, this list of integers is simplified further into a set of summarizing features.

\mypara{Feature representation.} To create a set of features for the WF classifier, some models \cite{DBLP:conf/uss/HayesD16} rely on manually engineered features that characterize network traces.  For instance, summary statistics about packet timing and direction (e.g., mean inter-arrival timing of packets, the number of consecutive packets sent in a given direction, etc.).  Differently, recent deep learning models automatically project a website access' trace representation into a latent feature space of lower dimensionality~\cite{DBLP:conf/ccs/SirinamIJW18} and then train over such features. However, none of the above models reason about the instants when monitored websites contained in multi-tab network traces are visited. In contrast, TSA-WF aims to better harness the information made available within trace representations to identify the approximate time at which monitored websites are visited in multi-tab samples.

\subsection{A Summary of WF Attacks \& Defenses}
\label{sec:existing-work}

We now summarize prominent WF attacks to better contextualize TSA-WF within this landscape, categorizing them based on the setting they address (i.e., whether they target the single- or the multi-tab setting). Then, we summarize existing WF defenses and briefly discuss their role in safeguarding users' privacy. 

\mypara{Single-tab setting.}  Early WF attacks, such as k-Fingerprinting~\cite{DBLP:conf/uss/HayesD16}, leverage manually crafted features combined with ML classifiers in the single-tab setting \cite{DBLP:conf/ccs/JuarezAADG14}. More recently, WF attacks focused on the extraction of (and reasoning about) latent features for performing single-tab WF attacks. Rimmer \textit{et al.}~\cite{DBLP:journals/corr/abs-1708-06376} evaluate the use of deep neural networks (DNNs) to automatically extract features and make single-tab classification decisions with high accuracy.  Deep Fingerprinting (DF)~\cite{DBLP:conf/ccs/SirinamIJW18} introduced a CNN-based architecture that uses a \textit{directional} representation, where traces are simply sequences of positive (+1) and negative (-1) ones, depending on whether a packet is incoming or outgoing. Tik-Tok~\cite{DBLP:journals/popets/RahmanSMG020} uses DF's architecture directly, while including packet timing information (by multiplying a packet's direction with its inter-arrival time).

\mypara{Multi-tab setting.} WF research has also focused on DL-based classifiers and feature extraction methods to expand attacks' scope towards a multi-tab setting. State-of-the-art DNNs have been pushing towards better accuracy and flexibility (e.g., number of website traces per sample, number of monitored websites) in multi-tab settings. ARES~\cite{DBLP:conf/sp/DengYLZLXXW23} uses an ensemble DNN to classify multi-tab traces with high accuracy, while Jin \textit{et al.}~\cite{DBLP:conf/ccs/JinLLS23} use a transformer model which can identify a potentially arbitrary number of websites within a multi-tab trace.

\mypara{WF defenses.} Among the most secure WF defenses, we find constant rate padding defenses, such as CS-BuFLO~\cite{DBLP:conf/wpes/CaiNJ14} and Tamaraw~\cite{DBLP:conf/ccs/CaiNWJG14}, which uniformize traffic by forcing the transmission of packets with a fixed size at a fixed rate. While these are able to successfully mask packet timing and burst characteristics, they incur high latency and bandwidth overheads. More efficient padding-based defenses include the adaptive padding (e.g., WTF-PAD~\cite{DBLP:conf/ccs/JuarezAADG14}) and randomized padding (e.g., FRONT~\cite{DBLP:conf/uss/GongW20}) approaches, which strategically transmit dummy packets to conceal the real network patterns generated by a given website access. A more recent defense that also aims to uniformize traffic, RegulaTor~\cite{holland2020regulator}, focuses on shaping the packet burst patterns that frequently occur in download traffic.
In our work, we concentrate on the padding-centric defenses discussed above, but refer the reader to an extensive analysis of the broader space of WF defenses~\cite{mathews2022sok}.

\section{WF as a Time Series Matching Problem}
\label{sec:three}

In this section, we introduce typical methods to visualize and reason about website traces as time series (\S\ref{sec:traces-as-timeseries}), discuss the challenges faced by earlier efforts for time series matching (\S\ref{sec:classical-ts-wf}), and justify the approach used in TSA-WF (\S\ref{sec:ts-matching-challenges}).

\subsection{Representing Traces as Time Series}
\label{sec:traces-as-timeseries}

\begin{figure}[t!]
    \centering
    \begin{minipage}{0.48\linewidth}
        \centering
        \includegraphics[width=\linewidth]{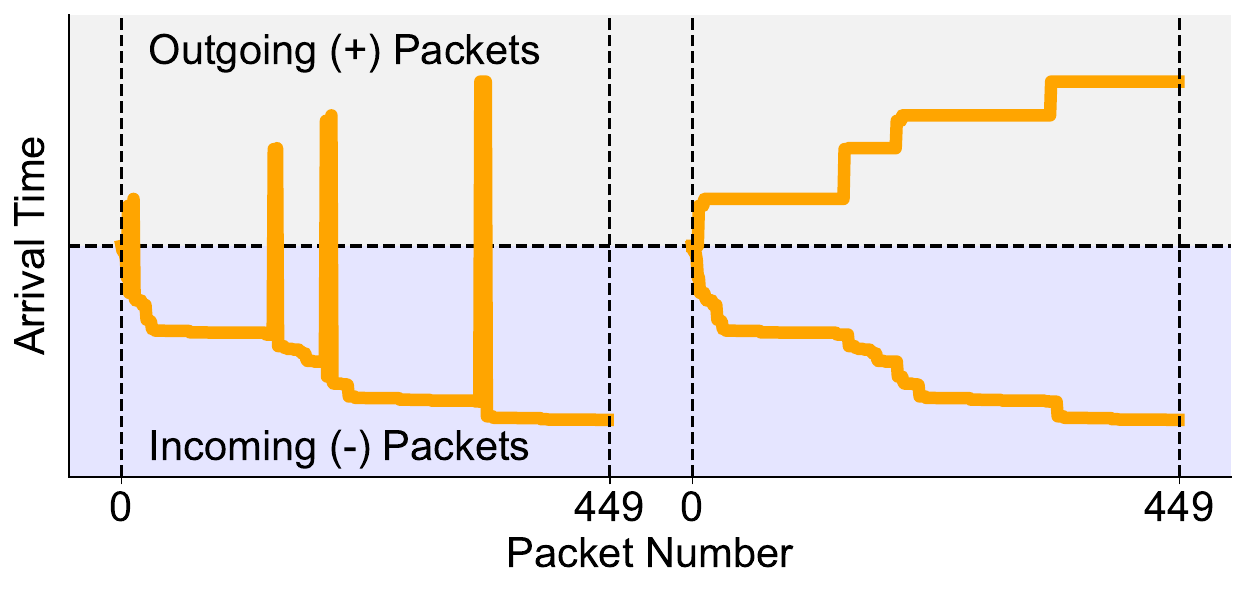}
        \vspace{-0.8cm}
        \caption{Trace with and without separation of incoming and outgoing packets.}
        \label{fig:single-split}
    \end{minipage}
    \hfill
    \begin{minipage}{0.48\linewidth}
        \centering
        \includegraphics[width=\linewidth]{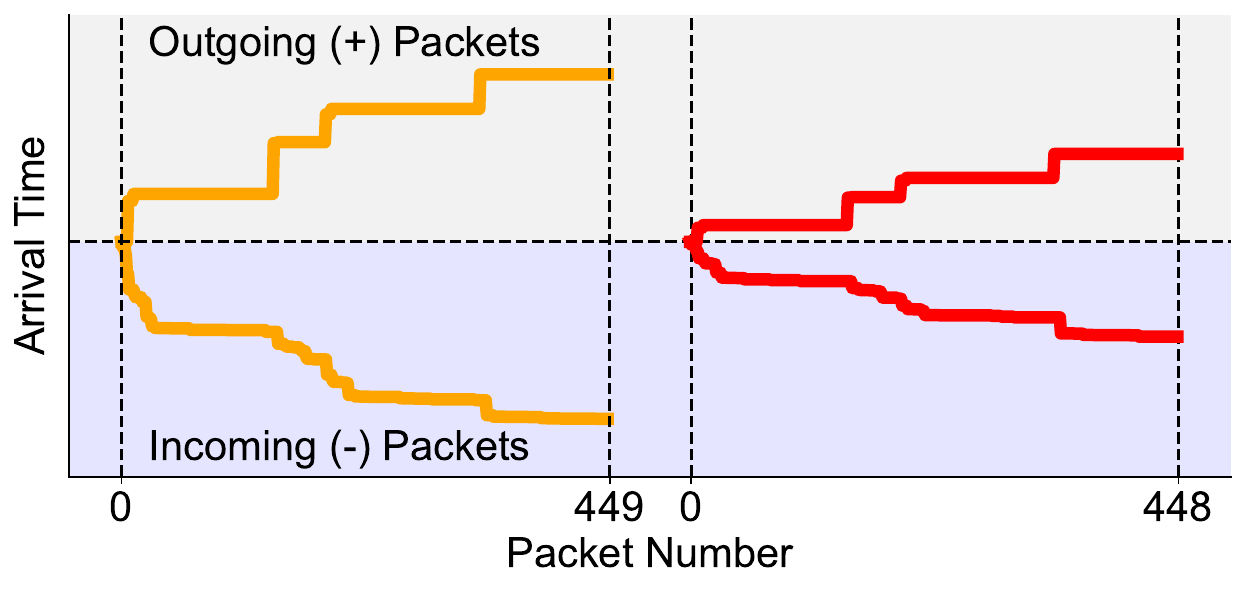}
        \vspace{-0.8cm}
        \caption{A time-series representation of two website traces from the same class.}
        \label{fig:same-class}
    \end{minipage}
    \vspace{0.2cm}
\end{figure}

In Fig. \ref{fig:single-split}, we depict two side-by-side representations of a website trace that an adversary might have obtained when observing (and successfully isolating) one of Alice's website accesses via Tor. We extract this trace from the Jin \textit{et al.}~\cite{DBLP:conf/ccs/JinLLS23} dataset, on which we expand further in \S\ref{sec:experiment}. This trace is composed by 449 packets and, for the purposes of this illustration, we ``reset'' the packet numbers shown in the \textit{x} axis to 0 in-between each representation. Below, we describe these representations and how they shape our techniques for comparing traces.

\mypara{Different representations are more adequate for matching.} The left-most website trace representation shown in Fig. \ref{fig:single-split} is arguably simple, obtained by directly visualizing packet traces as a time series where $x$ is the packet number, the sign of $y$ is the direction (positive for outgoing packets or negative for incoming ones), and the magnitude of $y$ is the arrival time. Since every packet is sorted by arrival time, $|y|$  increases for each subsequent packet within a trace.

In preliminary experiments we observed that, when algorithms based on time series similarity measures (e.g., euclidean distance) attempt to match time series represented in this format, they tend to produce unrepresentative similarity scores. This is due to two main factors: a) the representation makes for abrupt jumps in the value of $y$ when switching between incoming and outgoing packets, and; b) if there are horizontal delays (e.g., due to packet re-transmission) between two traces, this difference is reflected in both the positive and negative components of the time series. Existing work using time series analysis for WF has not addressed these issues, leveraging this simple representation~\cite{DBLP:conf/sp/RupprechtKHP19,DBLP:conf/wpes/NithyanandCJ14}.

To mitigate this issue, we propose a different representation of website traces, shown in the right-most section of Fig.~\ref{fig:single-split}, where we separately represent and compute the similarity between the outgoing and incoming components of a trace's time series (i.e., the similarity between any two traces is now given by two values which represent the traces' similarity in both the outgoing and incoming components). This operation tackles the issues identified above since between each timestamp, the difference in $y$ value, $d$, is always $d = |y_{i}|-|y_{i-1}|$, instead of $d = |y_{i}| + |y_{i-1}|$ should $x_{i-1}$ be outgoing while $x_{i}$ is incoming or vice-versa.

\mypara{Time series reveal intra-class variability.} Fig. \ref{fig:same-class} depicts the monitored website trace from Fig. \ref{fig:single-split} (in orange),  alongside another sample trace obtained from the same website in the single-tab setting (in red). While both traces appear reasonably distinct, we can observe that they roughly follow the same pattern with regards to the number, order, and arrival time of packets. In this example, we observe that the left-most trace exhibits a vertical shift of its pattern when compared to the right-most trace, which may be attributed to network delays that could have caused an initial delay w.r.t. the left-most trace. Such delays occur often, and are a well-known artifact of privacy-enhancing technologies such as Tor~\cite{DBLP:journals/pacmnet/NunesBCB0023}. Dealing with intra-class variability is a well-known challenge in WF attacks, and we discuss how TSA-WF can address it in \S\ref{sec:ts-matching-challenges}.

\begin{figure*}[t!]
    \centering
    \subfigure[Merged monitored (orange) and unmonitored (blue) traces.]{
        \includegraphics[width=0.45\textwidth]{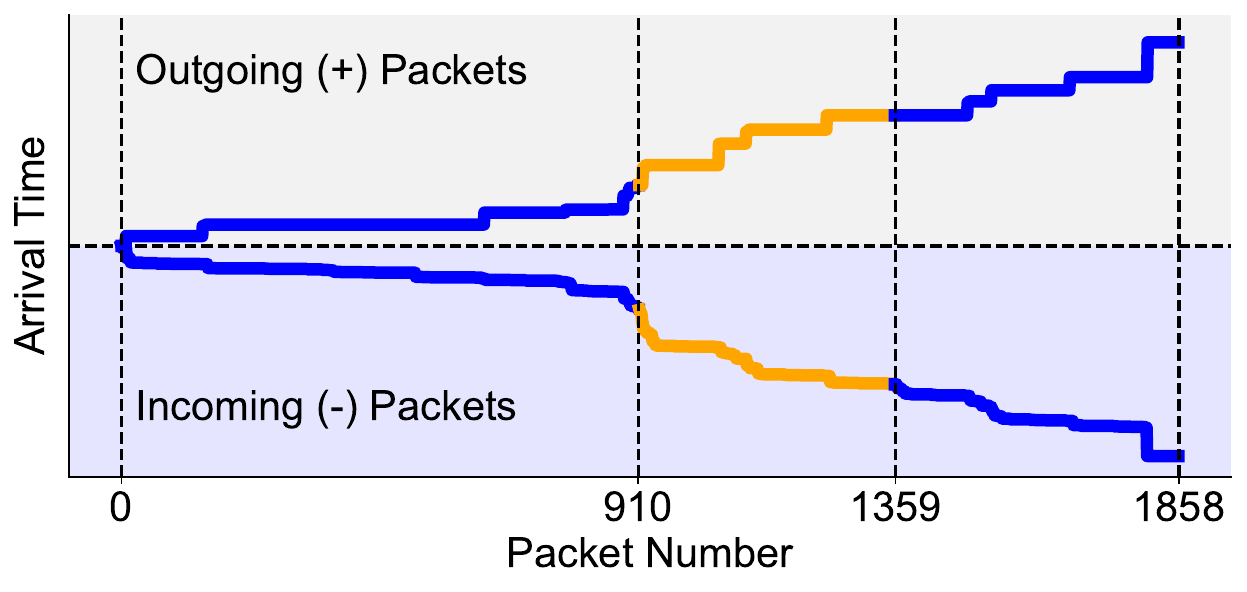}
        \label{fig:vert-merge}
    }
    \hfill
    \subfigure[Separated monitored (orange) and unmonitored (blue) traces.]{
        \includegraphics[width=0.45\textwidth]{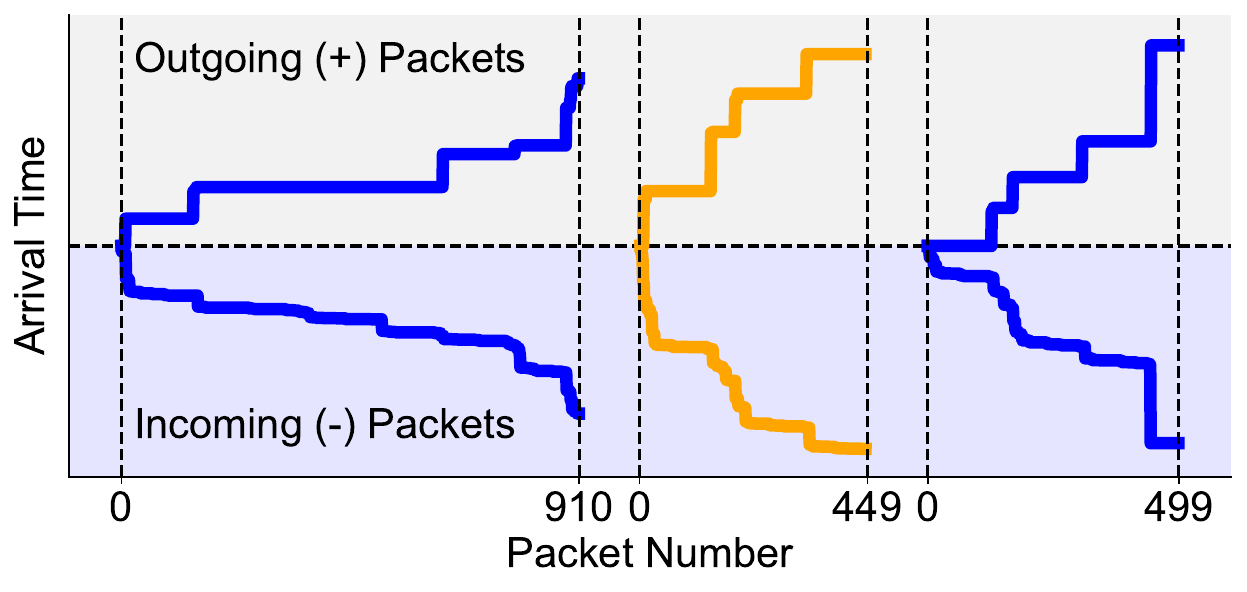}
        \label{fig:diff-class}
    }
    \vspace{-0.4cm}
    \caption{Merged and separated representations of monitored and unmonitored traces.}
    \label{fig:merged}
\end{figure*}

\mypara{Multi-tab traces are not easily separable.} Fig. \ref{fig:vert-merge} depicts a trace that an adversary might have recorded by sniffing Alice's network connection over a specific time period in the multi-tab setting. To better illustrate the components of Fig. \ref{fig:vert-merge}, assume that the packet subsequence that contains the monitored website from Fig. \ref{fig:single-split} is highlighted in orange, and that packet sequences that pertain to accesses to unmonitored websites are colored blue. These types of multi-tab traces make it hard for the adversary to reason about: a) how many websites have been visited over the considered time period; b) when does an access to a given website stop and the next one begin, and; c) how many websites from the monitored/unmonitored sets are included in the overall trace. As mentioned previously, past research efforts \cite{DBLP:conf/ccs/CuiCFCSC19,DBLP:conf/acsac/XuWLGCJ18,DBLP:journals/tdsc/YinLLWWSX22} have focused on parsing multi-tab traces and solving these questions automatically, but found limited success.

In Fig. \ref{fig:diff-class}, we show the results of an oracle which is able to cleanly separate the multi-trace into individual time series traces that reveal each of Alice's website accesses. TSA-WF aims to provide a method that can approximate the behavior of this oracle, thus separating multi-tab traces into its individual traces' components, and identify any monitored websites contained therein.

\subsection{Classical Time Series Matching for WF}
\label{sec:classical-ts-wf}

The representation of website traces as time series has been understood by researchers for a long time. In this section, we describe a set of WF attacks that do not fully harness the capabilities of existing time series matching techniques or use algorithms that appear to be suboptimal for the stated goals.

\mypara{Earlier time series analysis for WF.} To the best of our knowledge, dynamic time warping (DTW) has been the primary classical time series analysis technique that has seen a consistent use for matching website traces in the context of WF~\cite{DBLP:conf/sp/RupprechtKHP19,DBLP:journals/corr/abs-2110-01202,DBLP:conf/wpes/NithyanandCJ14}. DTW measures similarity between two time series that may vary in speed by aligning them non-linearly and stretching or compressing them to minimize the distance, making it robust to differences in length and temporal distortions. For instance, Rupprecht \textit{et al.} \cite{DBLP:conf/sp/RupprechtKHP19} used DTW to fingerprint websites accessed over LTE networks, while Nithyanand \textit{et al.}~\cite{DBLP:conf/wpes/NithyanandCJ14} leveraged DTW to build a WF defense where noise is added to packet traces (in the form of dummy packets) towards creating a large DTW distance between website traces. 

While DTW is simple and a natural choice for website traces, we show (\S\ref{sec:single-tab-experiment}) that it is not the best-performing time series analysis technique for WF datasets.

\mypara{Candidate time series analysis techniques for WF.} In this work, we explore the use of three additional time series matching techniques that are deemed effective for a wide range of applications which depend on the computation of time series' similarity, akin to the matching operations required for WF.

The first of such techniques is \textit{euclidean distance}, which simply measures the straight-line difference between corresponding points in two sequences, treating each point as a dimension in a multi-dimensional space. Euclidean distance is deemed simple, accurate, and effective but can be computationally intensive~\cite{middlehurst2024bake}.  This technique has been used in the past to detect attacks in network traffic~\cite{DBLP:journals/scn/ZhouWYZZ21}. The second technique we consider is \textit{STUMPY}~\cite{DBLP:journals/jossw/Law19}, which aims to make time series comparison fast and efficient by computing the euclidean distance between two time series using a matrix profile~\cite{DBLP:conf/icdm/YehZUBDDSMK16} along with further optimizations.  STUMPY has been previously used for time series anomaly detection~\cite{DBLP:journals/pvldb/HeLTWL23}. Finally, \textit{compression-based-distance} (CBD)~\cite{DBLP:journals/datamine/KeoghLRWLH07} is an indirect approach to comparing the distance between two time series.  In CBD, time series are converted into symbols (e.g. with symbolic aggregate approximation~\cite{DBLP:journals/algorithms/HeL0Z20}) that can be stored as a text file.  Then, these files are compressed (e.g., \texttt{.zip}, \texttt{.rar}) and the difference in their sizes are compared. Compression-based measures have been used before for anomalous web traffic analysis~\cite{DBLP:journals/igpl/Torre-AbaituaLA20}.

\subsection{Challenges for WF via Time Series Analysis}
\label{sec:ts-matching-challenges}

To better showcase the pitfalls of the time series analysis techniques introduced in the previous section, we consider an extension of the scenario from Fig.~\ref{fig:vert-merge}. In this example, the adversary uses a sample trace from its set of monitored websites (red trace from Fig.~\ref{fig:same-class}) and compares it with the multi-tab trace from Fig.~\ref{fig:vert-merge}.  The adversary aims to determine if and where the multi-tab trace contains an access to the same website of interest. Because the orange and red lines correspond to traces obtained from the same website, their distance \textit{should} be much smaller when compared to, e.g., the distance obtained when comparing the red line to the start of the multi-tab trace. Hence, in theory, an adversary should be able to find the location of the website trace which is in their monitored set by measuring the pattern similarity using \textit{euclidean distance}.

\begin{figure}[t!]
    \centering
    \vspace{-0.2cm}
    \includegraphics[width=0.52\textwidth]{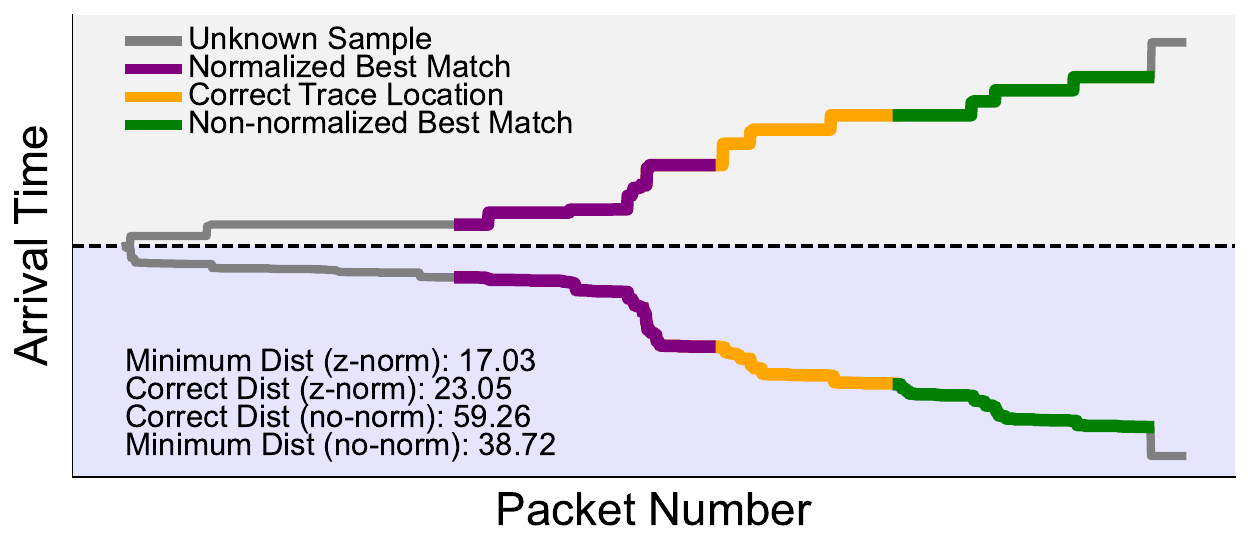}
    \vspace{-0.5cm}
    \caption{Best match locations for different techniques.}
    \label{fig:merged-location}
    \vspace{-0cm}
\end{figure}

Fig. \ref{fig:merged-location} shows how mismatches may occur when using both normalized and non-normalized euclidean distance (as approximated by STUMPY) to perform distance calculations. These calculations reveal that the sample website trace included in the adversary's monitored set was a better match with other sections of the multi-tab trace, which do not actually represent other samples from the monitored website of interest (note how the best matches in purple and green do not adequately overlap the orange line).  We will now showcase a set of challenges that may hamper one's ability to accurately compare website traces by solely resorting to euclidean distance, and propose mitigations for these mismatches.

\begin{figure*}[t!]
    \centering
    \subfigure[Random portions of a merged trace appear similar in normalized traces.]{
    \includegraphics[width=0.3\textwidth]{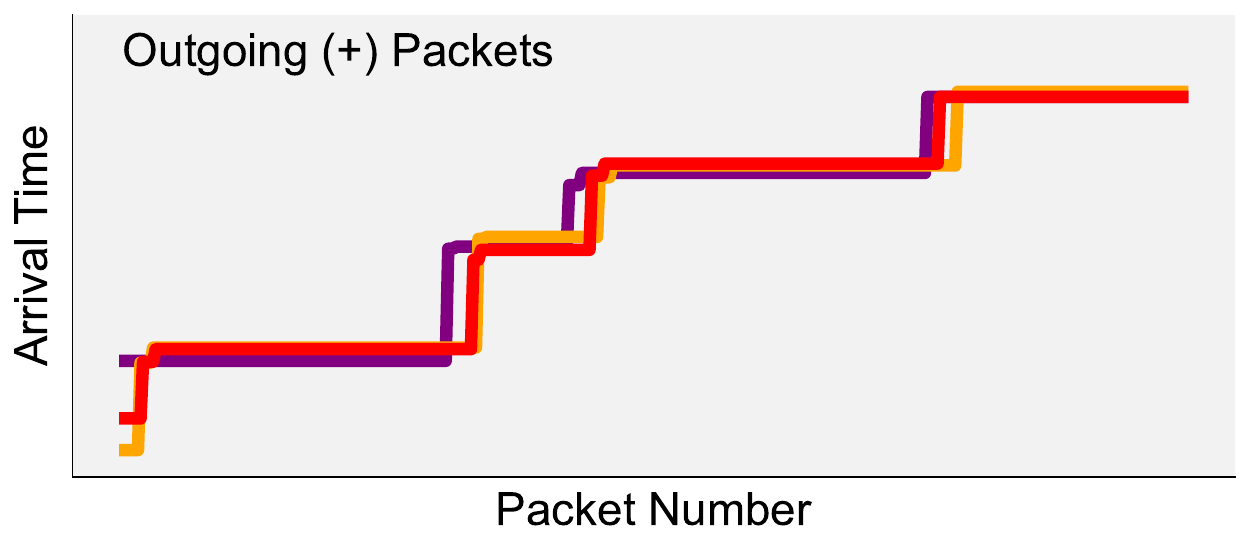}
        \label{fig:normalize}
    }
    ~\hfill
    \subfigure[Without normalization,  mismatched locations differ significantly.]{
    \includegraphics[width=0.3\textwidth]{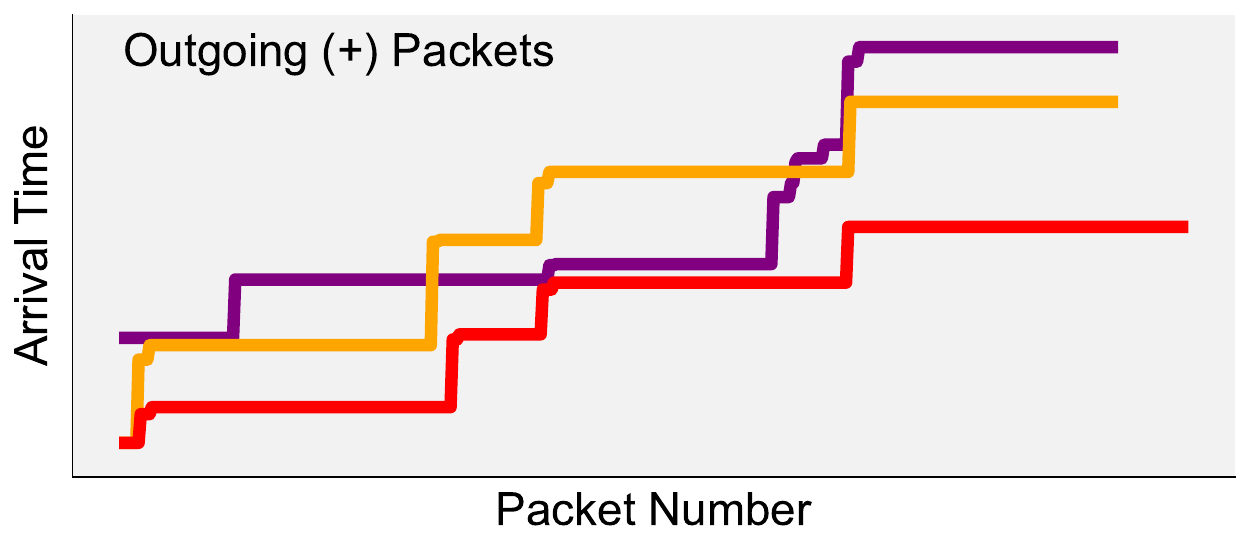}
        \label{fig:no-normalize}
    }
    ~\hfill
    \subfigure[Despite the visual mismatch, euclidean dist. at incorrect locations is smaller.]{
    \includegraphics[width=0.3\textwidth]{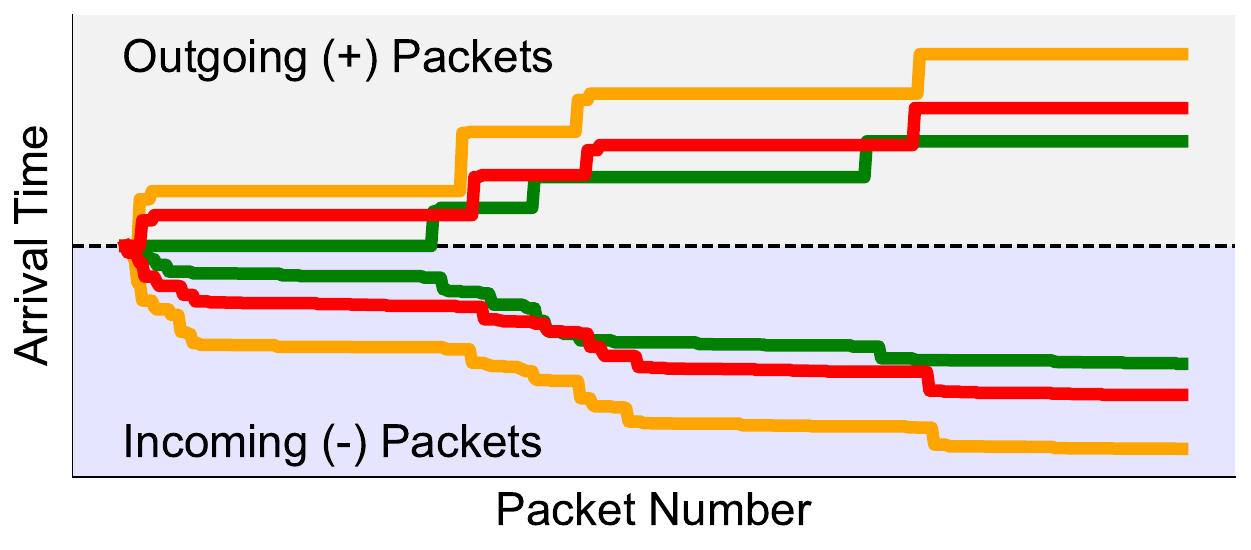}
        \label{fig:cbd}
    }
    \vspace{-0.3cm}
    \caption{Comparing the similarity of time series with and without normalization.}
    \label{fig:comparison}
\end{figure*}

\mypara{Concerns with normalized traces.} The first concern we address is tied to the fact that attempts to normalize traces before computing their similarity can lead to the homogenization of website traces which are very different in nature. In Fig.~\ref{fig:normalize} we overlay three website traces: the orange component of the merged trace from Fig.~\ref{fig:merged-location}, the red trace from Fig.~\ref{fig:same-class}, and the purple mismatched location from Fig.~\ref{fig:merged-location}.  Recall that we are attempting to find the location of the orange trace by comparing the time series distance between the merged trace from Fig.~\ref{fig:vert-merge} and the red trace from Fig.~\ref{fig:same-class}.  Note how the red line \textit{should} have a smaller euclidean distance to the orange compared with the purple. However, we show in Fig. ~\ref{fig:merged-location} that, after normalization, the distance between the red and orange lines is 23.05, but the distance between the red and purple lines is 17.03.  In contrast, we overlap the same traces (with the same colors) in Fig.~\ref{fig:no-normalize}, before applying any normalization. Observe that now the best match euclidean distance is found in a different location in Fig.~\ref{fig:no-normalize} with a distance of 38.72. Thus, we argue that normalized distance is inadequate for matching website trace representations, as it generates an apparent similarity when none actually exists.

\mypara{Concerns with non-normalized traces.} The second concern we address is that ignoring normalization can also lead to website traces with similarity scores that do not reflect their true nature (i.e., class).  In Fig.~\ref{fig:cbd} we overlap three traces: the orange component of the merged trace from Fig.~\ref{fig:merged-location}, the red trace from Fig. \ref{fig:same-class}, and the green mismatched location from Fig. \ref{fig:merged-location}. We also show in Fig. ~\ref{fig:merged-location} that, without normalization, the distance between the red and orange lines is 59.76, but the distance between the red and green lines is 38.72. Thus, large vertical shifts between two traces from the same website (e.g., caused by network delays, as in Fig.~\ref{fig:same-class}) can lead non-normalized euclidean distance-based matching to produce inaccurate results. While normalization can fix this issue, it introduces the aforementioned inconsistencies.

The two issues identified above with (non-)normalized euclidean distance call for the consideration of an additional time series comparison measure that can complement euclidean-based distance calculations. To this end, when implementing TSA-WF (\S\ref{sec:distance}), we explore the use of CBD distance calculations which eschew the magnitude of $y$ values (mitigating the non-normalization issue), without homogenizing different traces (mitigating the normalization issue).

\mypara{Concerns with proper subsequence weighting.} Lastly, when measuring euclidean distance, we can assign a particular weight to each point-wise pair of packets, reflecting the relative importance of this pair for the overall website trace distance. One observation about WF attacks is that the first few packets are far more important than the rest when issuing classification decisions~\cite{DBLP:conf/uss/GongW20}. Weights defined using an exponential decline (e.g., $w(x) = 0.5^x$) reflect this observation by maximizing the relative importance of initial packets. TSA-WF implements a variant of euclidean distance with exponentially declining weights, and our evaluation showcases the trade-offs of other weighting options (\S\ref{sec:parameters}).

\section{The Architecture of TSA-WF}
\label{sec:method}

Despite the challenges involved in performing accurate time series matching, the approaches detailed in \S\ref{sec:three} suggest that classical time series analysis techniques can be leveraged (and perhaps combined) to launch successful WF attacks. To validate this claim and assess practical limitations of this approach, we design and implement a WF pipeline named TSA-WF \textit{(\underline{T}ime-\underline{S}eries \underline{A}nalysis for \underline{W}ebsite \underline{F}ingerprinting)}. We implemented TSA-WF in $\sim$5\,000 lines of Python code to enable further experimentation by the WF research community~\cite{anon}.  
The dataset fed as input to TSA-WF consists of website traces obtained in either the single-tab or multi-tab setting stored as their trace representation. In our work, we assume that single-tab samples contain either a monitored or unmonitored trace, while multi-tab samples contain \textit{precisely one} monitored trace (see \S\ref{sec:goals-and-method}).

\begin{figure*}[t!]
    \centering
    \includegraphics[width=\linewidth]{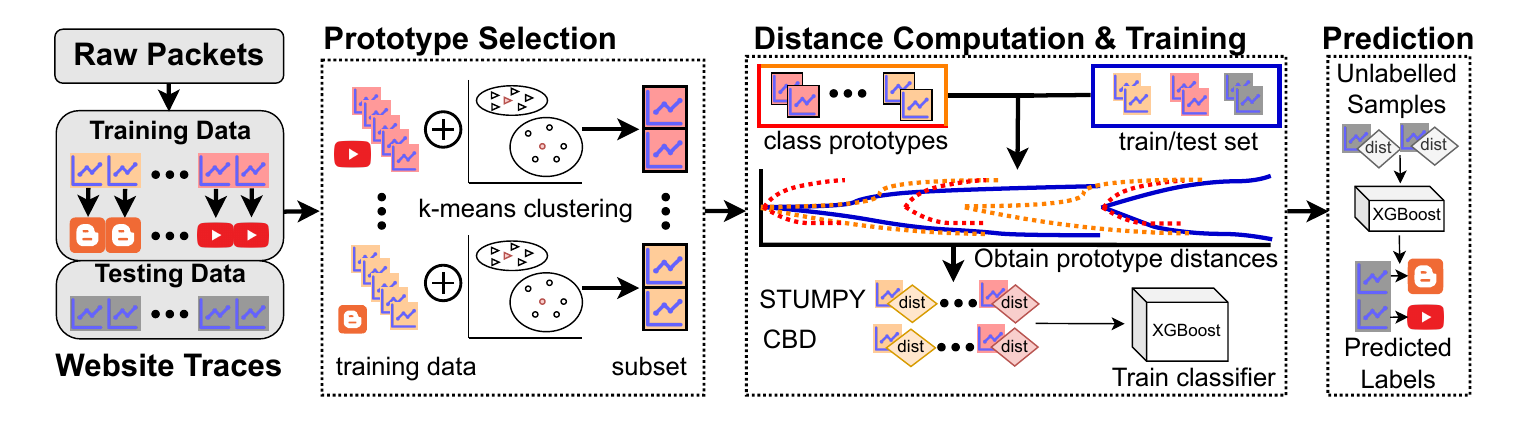}
    \vspace{-0.8cm}
    \caption{Depiction of TSA-WF's pipeline for performing a WF attack.}
    \label{fig:flowchart}
\end{figure*}

In TSA-WF's execution pipeline (shown in Fig.~\ref{fig:flowchart}), the adversary first obtains \textit{prototypes}, which are representative samples for each website in the monitored set. Then, during \textit{distance computation and training}, these prototypes are compared with samples from the training set using time series distance measures to train a classifier.  Finally, the classifier is evaluated against unlabeled samples based on its ability to \textit{predict} the monitored website and \textit{untangle the multi-tab trace}. Next, we detail each of these phases.

\subsection{Prototype Selection}
\label{sec:prototype-selection}

The first phase of the TSA-WF pipeline determines which samples to use as the best representatives of a website (i.e., a class' prototypes). For example, the class representative in our running example throughout \S\ref{sec:ts-matching-challenges} (see Fig. \ref{fig:same-class}) is the red trace the adversary uses to locate the orange monitored website. Below, we explore three different ways an adversary could select class prototypes.

\mypara{Random selection.} Prototypes could be chosen by randomly selecting a few traces per website from the training set. However, this may lead to choosing prototypes based on traces that a) might be outliers for a particular class, or that; b) are not sufficiently representative of diverse intra-class patterns. 


\mypara{Clustering raw traces.} In this approach, we organized each trace into a number of features equal to the number of packets. The value of each feature is equal to the packet's $y$-value in the time series (i.e., time and direction). To select prototypes, we applied k-means clustering to the feature set, and the centroid of each cluster is ultimately selected as a class prototype.
We observed that packet traces obtained by visiting the same website tend to form clearly defined clusters. While we hypothesize that these clusters may result from varying network conditions while traces are collected from the live Tor network~\cite{DBLP:journals/pacmnet/NunesBCB0023}, the above suggests that class prototypes can indeed be chosen using clustering.

\mypara{Clustering of trace features.} Despite producing visually separable clusters, the previous approach treats the integer representation of each packet as a separate feature for clustering. In this setting, k-means has to operate over potentially thousands of dimensions, leading to ``curse of dimensionality'' issues that statistical models suffer from when reasoning about samples in highly-dimensional spaces. As an alternative, we extracted the set of $\approx$150 summary statistics used by Hayes \textit{et al.} \cite{DBLP:conf/uss/HayesD16} from each website trace and used those features for clustering.  Again, the features are passed into k-means and the traces at the centroid are chosen to be the prototypes for each class. However, our experiments (\S\ref{sec:parameters}) suggest that this is only marginally more effective than clustering with raw traces.

\subsection{Distance Computation and Training}
\label{sec:distance}

The second phase of TSA-WF computes the distance between the (few) selected prototypes from each website (i.e., class) and all other trace samples in the training data. These distances will be later used to train a classifier assuming that the distances from a class' prototypes to samples of its own class should be smaller that the distances obtained when comparing the prototype to samples which belong to other classes. TSA-WF combines the distances computed with euclidean (with and without weights), STUMPY, CBD, and DTW by representing trace similarity as a vector $n$ with five cells (one for each distance measure we consider).

\mypara{Calculating distances.} Algorithm~\ref{alg:distance} details the computation of the distance between a class prototype and a training sample. Function \texttt{ComputeDist} (line~4) aims to find the minimum distance between a prototype $s$ and sample $t$ (both stored as lists containing the trace representation).  Specifically, \texttt{ComputeDist} finds the subsequence of $t$ that is most similar to $s$ (line~8) using each similarity function stored in $n$ (line~3) independently, then concatenates the final result. For every distance function in $n$ (line~6), we use a sliding window with size  $len(s)$ and step size 1 (line~7) that moves across $t$ and computes the distance at each interval to produce a matrix of distances $d$ with shape $[len(t) - len(s), len(n)]$ (line~5).  In the (unlikely) scenario that the prototype length exceeds the trace length, we reverse $t$ and $s$.  Then, TSA-WF reduces this matrix into a vector with shape $[n]$ by returning the minimum value of each distance measure (line~9).

\begin{algorithm}[t!]
\centering  
\caption{\scriptsize Compute best match between a prototype and a trace}
\label{alg:distance}
\resizebox{\linewidth}{!}{  
    \begin{minipage}{\textwidth}  
        \scriptsize  
        \begin{algorithmic}[1]
        \State $s \gets [...]$ \Comment{Prototype}
        \State $t \gets [...]$ \Comment{Trace}
        \State $n \gets [$STUMPY(), euclid(w=none), euclid(w=$e^{-x}$), CBD(), DTW()$]$ \Comment{Measures}
        \Procedure{ComputeDist}{s, t}
            \State $d \gets [len(t)-len(s),n]$ \Comment{Distances}
            \For{$j = 0 \rightarrow len(n)$, compute\_dist = n[j]} \Comment{Measures}
                \For{$i = 0 \rightarrow len(t), i \mathrel{+}= len(s)$} \Comment{Sliding window}
                    \State $d[i,j] =$ compute\_dist$(s, t[i:i+s])$ \Comment{Compute distance}
                \EndFor 
            \EndFor 
            \State \textbf{return} min($d$[:,$j$]) \Comment{Minimum of each distance measure}
        \EndProcedure
        \end{algorithmic}
    \end{minipage}
}
\end{algorithm}

\mypara{Training a classifier.} After the distances between all class prototypes and samples comprising the training set have been computed, TSA-WF trains a classifier using the resulting $[n]$-shaped distance vectors. 
A simple approach for using distance vectors to issue predictions is to determine a threshold for each class, such that every sample either belongs to a particular class or not depending on whether its distance is below a set threshold for that class. The thresholds for each class can be determined by recording the distance scores obtained by comparing a prototype with training samples from the same class. In our experiments, we report the best matching label for both single- and multi-tab traces.

A threshold classifier assigns labels based on the distance between the sample and other prototypes from the \textit{same} class.  However, it does not evaluate the distance between the sample and the prototypes of all possible classes.  By using an ML-based classifier, TSA-WF makes labeling decisions based on the similarity between the prototype and examples of every \textit{other} class as well.  Our results (\S\ref{sec:parameters}) suggest that gradient-boosted decision trees (as implemented by XGBoost)~\cite{DBLP:conf/kdd/ChenG16} make for an effective model for TSA-WF's final classification step.

\subsection{Prediction}
\label{sec:classification}

The final phase of TSA-WF predicts labels for a set of samples where the label associated with the monitored website contained therein is unknown. 
Recall that we assume unlabeled single-tab samples to either contain monitored or unmonitored websites, while multi-tab traces contain a single monitored website amongst an arbitrary number of unmonitored websites.  First, the distance between each unlabeled sample and the class prototypes used for training are calculated. Then, the classifier included in TSA-WF's pipeline (\S\ref{sec:distance}) assigns a label based on the training class that the sample is most similar to.

\subsection{Untangling Multi-Tab Traces}
\label{sec:improve-dl-method}

Given an unlabeled trace sample acquired when eavesdropping Alice's encrypted website accesses, an adversary interested in launching a multi-tab WF attack must make two decisions: a) which monitored website(s) are contained within the sample, and; b) at which instant is the monitored website located in that sample. All WF attacks must achieve the former by definition but, to the best of our knowledge, existing multi-tab attacks do not attempt the latter.  TSA-WF is capable of jointly performing a) and b), or use the results of another attack for a) and then compute b) independently. 

We determine the location of the monitored website within an unlabeled trace sample is as follows. First, we train a classifier that determines the class of the monitored trace within an unlabeled sample. This can either be TSA-WF's own classifier (described previously) or sourced from elsewhere (e.g., a pre-existing WF attack). Next, we record the labels assigned by the classifier for each unlabeled sample. Finally, we select one prototype from the class assigned by the chosen WF attack and compute the distance between it and the trace sample. The packet index where the smallest distance was computed is TSA-WF's guess for the location of the monitored website within a multi-tab sample.

\section{Evaluation} 
\label{sec:experiment}

We now describe our experimental setup (\S\ref{sec:goals-and-method}) and how we tuned TSA-WF's parameters (\S\ref{sec:parameters}).  Then, we evaluate TSA-WF on single- (\S\ref{sec:single-tab-experiment}) and multi-tab (\S\ref{sec:multi-tab-experiment}) settings, and its ability to identify the instant at which a target website is accessed (\S\ref{sec:improve-dl}). Lastly, we evaluate TSA-WF's computational efficiency (\S\ref{sec:compefficiency}).

\subsection{Experimental Setup}
\label{sec:goals-and-method}

\mypara{Dataset.} We evaluate TSA-WF with the Tor traffic dataset used in the state-of-the-art multi-tab DL-based attack of Jin \textit{et al.}~\cite{DBLP:conf/ccs/JinLLS23}. 
It contains 100 traces per each of 50 different monitored websites ($100 * 50 = 5\,000$) and 5\,000 traces from unmonitored websites, totaling 10\,000 single-tab traces.  
We generate 1\,000 multi-tab traces for each monitored website ($1\,000 * 50 = 50\,000$), and divide them into a 90/10 training/testing split. To create a multi-tab trace with $x$ tabs, we take one random monitored website and merge/overlap it with $x-1$ random unmonitored website traces. Thus, each multi-tab trace contains precisely one monitored website and, in each experiment, a WF attack must identify which of the 50 monitored websites was included in the trace. This is consistent with prominent multi-tab WF attacks (e.g.,~\cite{DBLP:conf/sp/DengYLZLXXW23}) that also finely control the composition of multi-tab traces. We experiment with simply merged traces (i.e., overlap = 0\%), as well as three different settings where we assume that each trace comprising a merged trace may overlap adjacent traces up to 10\%, 20\%, or 40\%.

In line with existing WF attacks~\cite{DBLP:conf/uss/HayesD16,DBLP:conf/ccs/SirinamIJW18,DBLP:journals/popets/RahmanSMG020,DBLP:conf/sp/DengYLZLXXW23} against which we compare TSA-WF to, we gauge attacks and defenses on a \textit{synthetic} dataset where website traces are obtained by using automated browsers to crawl popular URL lists over Tor~\cite{DBLP:conf/ccs/JinLLS23}. While Cherubin et al.~\cite{DBLP:conf/uss/CherubinJT22} found that experiments using synthetic WF datasets can overestimate attack performance, these datasets are still widely used~\cite{10.1145/3658644.3670272,DBLP:conf/ccs/JinLLS23,DBLP:journals/popets/VeichtRB23} for evaluating both the performance of novel WF attacks and for determining the robustness of new defenses. Prior work has also shown that WF attacks tested on synthetic datasets can be adapted to real-world data by incorporating binary classifiers or multi-classification sub-tasks~\cite{DBLP:journals/tifs/WangXGQR22,DBLP:journals/popets/JansenW23}. We defer the incorporation of such adaptation techniques in TSA-WF to future work.



\mypara{Trace distance measures.} To compute the distance between websites' prototypes and other traces contained in TSA-WF's training/testing data, we use existing code for STUMPY, the \texttt{numpy} implementation of euclidean distance, the \texttt{tslearn} implementation for DTW, and our own implementation of CBD.

\mypara{WF attacks.} We compare TSA-WF in the single-tab setting against an influential ML-based WF attack, k-Fingerprinting~\cite{DBLP:conf/uss/HayesD16} (k-FP), as well as two DL-based attacks extensively used to benchmark WF attacks -- Deep Fingerprinting~\cite{DBLP:conf/ccs/SirinamIJW18} (DF) and Tik-Tok~\cite{DBLP:journals/popets/RahmanSMG020}. We compare the effectiveness of TSA-WF in the multi-tab setting with ARES~\cite{DBLP:conf/sp/DengYLZLXXW23}, a state-of-the-art multi-tab DL-based WF attack.

\mypara{Evaluation of attack performance.} To evaluate each WF attack, we report the classification accuracy, based on the ability of the WF attack to identify the correct label for the monitored website contained in a single-/multi-tab sample. Note that our accuracy measurement accounts for determining exact class labels, unlike the original open-world evaluation of DF~\cite{DBLP:conf/ccs/SirinamIJW18}, which only reports whether a given single-tab trace belongs to the monitored set or not.

\mypara{WF defenses.} We evaluate the considered WF attacks against traces shielded using three well-known WF defenses seen in \S\ref{sec:existing-work}: WTF-PAD~\cite{DBLP:journals/corr/JuarezIPDW15}, FRONT~\cite{DBLP:conf/uss/GongW20}, and RegulaTor~\cite{holland2020regulator}. In line with existing work~\cite{DBLP:journals/popets/VeichtRB23}, we use high-fidelity simulators from these defenses (made available by their authors) to generate defended traces based on the pre-recorded undefended traces contained in the Jin \textit{et al.}~\cite{DBLP:conf/ccs/JinLLS23} dataset. We used two configurations of FRONT (T1 and T2) as configured by Veicht \textit{et al.}~\cite{DBLP:journals/popets/VeichtRB23}. Due to its larger sampling window, {FRONT-T2} induces more dummy packets in the trace when compared to {FRONT-T1}. When generating multi-tab traces, we first apply the defense on each individual trace before merging/overlapping them. This provides an explicit advantage to FRONT since it adds dummy packets to the start of each trace.

\mypara{Laboratory testbed.} 
Most of our experiments were conducted on a MacBook Pro with an M1 Max CPU and 16GB of RAM. We used this machine to train TSA-WF in the single- and multi-tab settings, as well as the single-tab WF attacks we compare to. For experimenting with ARES, we used a Linux machine with an Intel Xeon E5-2650 CPU, 251GB RAM and 2 Nvidia Tesla P100 GPUs.

\subsection{Parameter Tuning}
\label{sec:parameters}

In this initial round of experiments, we limited data availability to samples pertaining to the simpler closed-world scenario and single-tab setting. Our task was to determine reasonable settings for five main parameters of TSA-WF: a) the method for selecting prototypes; b) the amount of prototypes;  c) the weight configuration for euclidean distance, and; d) the classification model.

When performing these experiments, we used STUMPY as our default distance measure (except when configuring weights for an actual implementation of euclidean distance) and a random forest classifier (except when testing alternative models to use in the final prediction step of TSA-WF). We experiment (and set) each of the aforementioned parameters in a cascade fashion.

\mypara{Clustering algorithm.} Table~\ref{table:clustering} shows the accuracy of TSA-WF when using different clustering algorithms for prototype selection. We evaluated four different tried-and-tested clustering algorithms: \textit{k}-means, Affinity Propagation (AP), DBSCAN and OPTICS. 
In this experiment, we clustered the traces belonging to each monitored website in the training set (using the summary statistics used as part of k-Fingerprinting~\cite{DBLP:conf/uss/HayesD16} -- see \S\ref{sec:prototype-selection}), and selected the resulting centroids as each website's prototypes (to a maximum of two). We can see from the table that the default TSA-WF configuration using \textit{k}-means (generating two clusters) outperforms TSA-WF setups using the remaining three methods, achieving an accuracy of 88.8\%. We select this clustering algorithm for protoype selection.

\begin{table}[t!]
    \centering
    \begin{minipage}{0.28\linewidth}
        \centering
        \vspace{-0.1cm}
        \caption{Accuracy of the clustering methods.}
        \label{table:clustering}
        \vspace{-0.2cm}
        \resizebox{0.90\linewidth}{!}{
        \begin{tabular}{@{}lc@{}}
            \toprule
            \textbf{Method} & \textbf{Accuracy} \\ 
            \midrule
            \textit{k}-means & 0.888 \\ 
            AP & 0.885 \\ 
            DBSCAN & 0.872 \\ 
            OPTICS & 0.873 \\ 
            \bottomrule
        \end{tabular}
        }
    \end{minipage}
    \hfill
    \begin{minipage}{0.38\linewidth}
        \centering
        \vspace{-0.1cm}
        \caption{Accuracy of the prototype clustering features.}
        \label{table:prototype}
        \vspace{-0.2cm}
        \resizebox{\linewidth}{!}{
        \begin{tabular}{@{}clll@{}}
            \toprule
            \textbf{\# Protos.} & \textbf{Random} & \textbf{Clusters} & \textbf{k-FP} \\ 
            \midrule
            \textbf{1} & 0.864 & 0.870 & 0.870 \\
            \textbf{2} & 0.871 & 0.880 & \textbf{0.888} \\
            \textbf{3} & 0.878 & 0.889 & 0.895 \\
            \textbf{4} & 0.883 & 0.893 & 0.900 \\
            \textbf{5} & 0.884 & 0.895 & \textbf{0.904} \\ 
            \bottomrule
        \end{tabular}
        }
    \end{minipage}
    \hfill
    \begin{minipage}{0.28\linewidth}
        \centering
        \vspace{-0.1cm}
        \caption{Accuracy of the ML-DL classifiers.}
        \label{table:classifier}
        \vspace{-0.2cm}
        \resizebox{0.955\linewidth}{!}{
        \begin{tabular}{@{}ll@{}}
            \toprule
            \textbf{Classifier} & \textbf{Accuracy} \\ 
            \midrule
            Decision Tree & 0.860 \\
            Random Forest & 0.888 \\
            XGBoost & \textbf{0.890} \\
            CNN & 0.790 \\
            MalConv & 0.830 \\ 
            \bottomrule
        \end{tabular}
        }
    \end{minipage}
    \vspace{-0cm}
\end{table}

\mypara{Clustering features.} Next, in Table \ref{table:prototype}, we compared the performance of three different prototype selection methods, when applied to \textit{k}-means: a) random selection; b) clustering with the raw trace representation, and; c) clustering with k-FP's feature representation. Each technique was evaluated using 1 to 5 prototypes (i.e., \textit{k}-means' cluster centroids) per website. Overall, TSA-WF with 5 prototypes as selected via k-FP performed the best, attaining a maximum accuracy of 90.4\%, and outperforming random choice (88.4\%) and trace clustering (89.5\%). We note that the computational complexity of TSA-WF increases substantially with each additional prototype, while accuracy benefits are negligible. Thus, we select two prototypes from the pool of training samples for each class.

\mypara{Configuring the weighted euclidean distance.} As mentioned in \S\ref{sec:ts-matching-challenges}, WF research has found that the initial segment of each website trace carries the most valuable information for launching successful WF attacks. One of the distance metrics we experiment with as part of TSA-WF is a weighted euclidean distance (WED) scheme, wherein different segments of the trace are assigned varying weights depending on their relative position within the trace.  We investigated three weight-assigning strategies to get the weight vector $W = [w_1,w_2, \dots w_l]$: 

\begin{compactitem}
    \item \textit{Linear weights:} Computed as the inverse of the packet's position $k$: $W = [w_k =( l - k) / \sum_{i= 1}^l i]$ for $  1 \leq k \leq l$.
    \item \textit{Logarithmic weights}: Computed as a $log$ function: $W^{'} = [w_k = log(l-k)] $for $  1 \leq k \leq l$. The weight vector is normalized as: $W = W^{'} / \sum W^{'} $.
    \item \textit{Reflected logarithmic weights}: 
    Computed as $W^{'} = [w_k = log(k/\sum_{i= 1}^l i)]$ for $1 \leq k \leq l$, and $W = W^{'} / \sum W^{'} $.
\end{compactitem}
When evaluating TSA-WF using these variations of WED in lieu of STUMPY, we obtained an accuracy of 83.5\% for linear weights, 84.8\% for logarithmic weights, and  85.1\% for reflected logarithmic weights.  While WED does not excel as an independent distance measure when compared to STUMPY, its combination with other measures can enhance TSA-WF's accuracy (see \S\ref{sec:single-tab-experiment}). For the remainder of our experiments, we report WED using reflected logarithmic weights.

\mypara{Classifier choice.} In Table~\ref{table:classifier}, we show the performance of TSA-WF when paired with different classifiers.  We evaluate a pool of popular ML algorithms, including a decision tree, random forest, and XGBoost~\cite{DBLP:conf/kdd/ChenG16}, together with two neural network architectures based on a CNN and MalConv~\cite{DBLP:conf/aaai/RaffBSBCN18}. We can see that the use of XGBoost allowed TSA-WF to achieve the highest accuracy (89\%). From this point on, we incorporate XGBoost into TSA-WF for our subsequent experiments.  With all parameters now configured, the next sections evaluate TSA-WF in the more realistic open-world scenario.

\subsection{TSA-WF's Effectiveness on Single-Tab Traces}
\label{sec:single-tab-experiment}

In Table~\ref{tab:main-result}, we evaluate the efficacy of TSA-WF in the open-world scenario and single-tab setting (see the \texttt{1-Tab} column) and compare it to WF attacks that operate in the same regime.

\mypara{Classification using individual distance measures.} Initially, we evaluate our time-series method by employing each distance measure independently. STUMPY exhibits the highest classification accuracy (87.9\%) when compared with all other time series analysis methods (i.e., euclidean, WED, CBD, and DTW).

\mypara{Classification using a combination of distance measures.} As discussed in \S\ref{sec:ts-matching-challenges}, each similarity measure captures different aspects of website trace matching and confers benefits in certain respects while lacking in others. As anticipated, TSA-WF's accuracy increased when the pipeline's prediction model is supplied with the feature vectors obtained from multiple measure combined. Indeed, combining STUMPY, euclidean, WED, CBD and DTW, results in an accuracy of 92.2\% -- offering an improvement of 4.3\% over the use of STUMPY alone.

\mypara{Benchmarking WF attacks on undefended traces.} We also benchmarked a set of prominent WF attacks on the same dataset used to evaluate TSA-WF above. As shown in Table~\ref{tab:main-result}, TSA-WF's classification pipeline (fueled by all distance measures) outperformed Tik-Tok, DF, and k-FP, distancing itself from the accuracy achieved by k-FP by 2.4\%. 

\mypara{Benchmarking WF attacks on defended traces.} In Table~\ref{tab:single-tab-defense}, we compare TSA-WF's performance (using STUMPY) with other attacks in the open-world setting, when WF defenses were applied to traces. TSA-WF achieves similar performance to existing WF attacks, outperforming Tik-Tok by 1.5\% vs. {FRONT-T1} and 4.0\% vs. {FRONT-T2}, as well as outperforming k-FP by 2.7\% vs. RegulaTor. However, Tik-Tok outperforms TSA-WF by 0.8\% vs. WTF-PAD.

\begin{table}[t!]
\centering
\caption{Accuracy of different WF attacks in the single- and multi-tab (merged traces only) open-world scenario. Each column shows the scores using traces with 1 to 7 tabs.}
\label{tab:main-result}
\vspace{-0.2cm}
\resizebox{0.85\linewidth}{!}{
\begin{tabular}{@{}lcccc@{}}
\toprule
\multirow{2}{*}{\textbf{Attack Method}} & \multicolumn{4}{c}{\textbf{\# of Tabs}} \\ \cmidrule(lr){2-5}
                        & \textbf{1-Tab} & \textbf{3-Tab} & \textbf{5-Tab} & \textbf{7-Tab} \\ \midrule
STUMPY                  & 0.879          & 0.505          & 0.397          & 0.340          \\
Euclidean               & 0.816          & 0.440          & 0.340          & 0.298          \\
WED           & 0.875          & 0.480          & 0.369          & 0.380          \\
CBD                     & 0.758          & 0.384          & 0.260          & 0.204          \\ 
DTW                     & 0.857          & -              & -              & -              \\ \midrule
STUMPY + CBD            & 0.886          & 0.502          & 0.400          & 0.339          \\
STUMPY + Euclidean      & 0.897          & 0.517          & 0.400          & 0.346          \\
STUMPY + WED  & 0.909          & 0.517          & 0.408          & 0.344          \\
STUMPY + CBD + Euclidean& 0.893          & 0.520          & 0.406          & 0.345          \\
STUMPY + WED + Euclidean & 0.912  & 0.520          & 0.410          & 0.348          \\
STUMPY + WED + DTW      & 0.904  & -              & -              & -              \\
STUMPY + Euclidean + DTW           & 0.913  & -              & -              & -              \\
STUMPY + WED + Euclidean + CBD & 0.913 & \textbf{0.522} & \textbf{0.412} & \textbf{0.351} \\ 
STUMPY + WED + Euclidean + DTW & 0.915  & -              & -              & -              \\ 
All                     & \textbf{0.922} & -              & -              & -              \\ \midrule
ARES~\cite{DBLP:conf/sp/DengYLZLXXW23} & - & 0.875 & 0.710 & - \\ \midrule
DF~\cite{DBLP:conf/ccs/SirinamIJW18} & 0.758 & - & - & - \\
Tik-Tok~\cite{DBLP:journals/popets/RahmanSMG020} & 0.796
 & - & - & - \\
k-FP~\cite{DBLP:conf/uss/HayesD16} & 0.898 & - & - & - \\ \bottomrule
\end{tabular}
}
\end{table}

\begin{table}[t]
\centering
\caption{Accuracy of WF attacks in defended single-tab open-world traces.}
\vspace{-0.2cm}
\label{tab:single-tab-defense}
\resizebox{0.85\linewidth}{!}{
\begin{tabular}{@{}lcccc@{}}
\toprule
\multirow{2}{*}{\textbf{Attack}} & \multicolumn{4}{c}{\textbf{Defenses}} \\ \cmidrule(l){2-5}
 & \textbf{WTF-PAD} & \textbf{FRONT-T1} & \textbf{FRONT-T2} & \textbf{RegulaTor} \\ \midrule
TSA-WF \footnotesize{(STUMPY)} & 0.735 & \textbf{0.722} & \textbf{0.756} & \textbf{0.633} \\
DF & 0.727 & 0.664 & 0.677 & 0.490 \\
Tik-Tok & \textbf{0.743} & 0.707 & 0.716 & 0.508 \\
k-FP & 0.692 & 0.656 & 0.709 & 0.606 \\ \bottomrule
\end{tabular}
}
\end{table}

\subsection{TSA-WF's Effectiveness on Multi-Tab Traces}
\label{sec:multi-tab-experiment}

\mypara{Merged traces.} In this experiment, we compare the performance of TSA-WF with existing work designed for the multi-tab setting in the open-world scenario. First, similarly to our observations in the single-tab setting, we can see in Table~\ref{tab:main-result} that the standalone use of STUMPY in TSA-WF's pipeline still reaps most of the benefits when compared to the combined usage of distance measures. Specifically, we can observe that TSA-WF (STUMPY) achieves an accuracy of 50.5\% for the \texttt{3-Tab} scenario, while the combination of STUMPY, WED, euclidean, and CBD distances achieves an accuracy of 52.2\%, only 1.7\% better. However, we also see that, in the multi-tab setting, CBD helped improve TSA-WF's accuracy, as opposed to the single-tab setting where its use provided no advantage. 

Second, we see that TSA-WF was not able to outperform state-of-the-art deep learning-based attacks designed for the multi-tab setting. The table shows that TSA-WF (equipped with STUMPY only) achieved an accuracy of 50.5\% in the \texttt{3-Tab} setting and 39.7\% in the \texttt{5-Tab} setting, revealing a gap of 37\% and 31.3\% in accuracy, respectively, when compared with ARES. 

Note that we did not evaluate ARES in the \texttt{7-Tab} setting as it truncates traces after 10k packets (similarly to DF's and Tik-Tok's behavior), thus losing important contextual information---possibly the entirety of some website traces contained within a given multi-tab trace. While it would be possible to admit larger inputs by adding hidden layers to ARES (e.g., additional gated recurrent units), this would require non-trivial transformations to ARES's DNN. Nevertheless, with such changes, we anticipate that ARES would continue to outperform TSA-WF in this setting. We also did not evaluate DTW on multi-tab traces since distance calculations take much longer (see \S\ref{sec:compefficiency}), making it intractable.

\begin{table}[t]
    \centering
    \begin{minipage}{0.48\linewidth}
        \centering
        \caption{Acc. of TSA-WF and ARES in defended multi-tab open-world traces.}
        \label{tab:multi-tab-defense}
        \vspace{-0.2cm}
        \resizebox{0.9\linewidth}{!}{
        \scriptsize
        \begin{tabular}{@{}llccc@{}}
        \toprule
        \multirow{2}{*}{\textbf{Attack}} & \multirow{2}{*}{\textbf{Defense}} & \multicolumn{3}{c}{\textbf{\# of Tabs}} \\ \cmidrule(l){3-5}
        &  & \textbf{3-Tab} & \textbf{5-Tab} & \textbf{7-Tab} \\ \midrule
        & WTF-PAD & 0.278 & 0.185 & 0.146 \\
        TSA-WF & FRONT-T1 & 0.295 & 0.213 & 0.165 \\
        (STUMPY) & FRONT-T2 & 0.313 & 0.220 & 0.176 \\
        & RegulaTor & \textbf{0.280} & \textbf{0.208} & 0.171 \\ \midrule
        & WTF-PAD & \textbf{0.584} & \textbf{0.416} & - \\
        ARES & FRONT-T1 & \textbf{0.469} & \textbf{0.324} & - \\
        & FRONT-T2 & \textbf{0.581} & \textbf{0.391} & - \\
        & RegulaTor & 0.071 & 0.097 & - \\ \bottomrule
        \end{tabular}
        }
    \end{minipage}
    \hfill
    \begin{minipage}{0.48\linewidth}
        \centering
        \caption{Acc. of TSA-WF and ARES in the multi-tab open-world scenario, where traces overlap between 0\% and 40\%.}
        \label{tab:overlaps}
        \vspace{-0.2cm}
        \resizebox{0.9\linewidth}{!}{
        \scriptsize
        \begin{tabular}{@{}lccccc@{}}
        \toprule
        \multirow{2}{*}{\textbf{Attack}} & \multirow{2}{*}{\textbf{Overlap}} & \multicolumn{3}{c}{\textbf{\# of Tabs}} \\ \cmidrule(lr){3-5}
        & & \textbf{3-Tab} & \textbf{5-Tab} & \textbf{7-Tab} \\ \midrule
        \multirow{3}{*}{TSA-WF}
        & 10\% & 0.450 & 0.344 & 0.295 \\
        & 20\% & 0.423 & 0.314 & 0.254 \\
        (STUMPY) & 40\% & 0.386 & 0.273 & 0.210 \\ \midrule
        \multirow{3}{*}{ARES} 
        & 10\% & 0.806 & 0.619 & - \\
        & 20\% & 0.771 & 0.590 & - \\
        & 40\% & 0.761 & 0.586 & - \\ \bottomrule
        \end{tabular}
        }
    \end{minipage}
\end{table}

\mypara{Defended merged traces.} Table~\ref{tab:multi-tab-defense} describes the accuracy of TSA-WF (w/ STUMPY) on defended traces created from 3/5/7 tabs. 
Recall TSA-WF's undefended accuracies of 50.5\% and 39.7\% in the \texttt{3-Tab} and \texttt{5-Tab} settings.  The results of Table~\ref{tab:multi-tab-defense} show a decrease in accuracy between 19.2\% (vs. {FRONT-T2}) and 22.7\% (vs. WTF-PAD) for \texttt{3-Tab} and between 17.7\% (vs. {FRONT-T2}) and 21.2\% (vs. WTF-PAD) for \texttt{5-Tab}. Recall ARES' undefended accuracy of 87.5\% and 71.0\% in the \texttt{3-Tab} and \texttt{5-Tab} settings. Table~\ref{tab:multi-tab-defense} shows ARES experienced an accuracy reduction between 29.1\% (vs. WTF-PAD) and 80.4\% (vs. RegulaTor) for \texttt{3-Tab} and between 29.4\% (vs. WTF-PAD) and 65.6\% (vs. RegulaTor) for \texttt{5-Tab}.  While RegulaTor was the most effective defense against TSA-WF for \texttt{1-Tab} (see Table~\ref{tab:single-tab-defense}), WTF-PAD was the best defense against it for 3 and 5 tabs.  Overall, TSA-WF had a smaller relative decrease in accuracy compared with ARES, but still failed to outperform ARES against most defenses.

We note that ARES is highly ineffective against traces defended using RegulaTor and posit two reasons for this result. Firstly, the deep learning architecture of ARES is closely related to that of DF -- indeed, ARES' local profiling component uses the same 32-layer CNN architecture as DF. For this reason, we expect ARES' performance to be similarly impacted by the same perturbations introduced by the RegulaTor defense.Secondly, RegulaTor traces are significantly longer compared to the other defenses.  On average, when considering the \texttt{3-Tab} setting, ARES truncates RegulaTor-generated traces at 46\% of their length, vs. only $\approx$30\% for {FRONT-T1}, {FRONT-T2}, and WTF-PAD.  Trace truncation is likely to have a negative impact on the performance of ARES, although we saw similarly low scores in both the \texttt{3-Tab} and \texttt{5-Tab} settings. We note that ARES has not previously been evaluated against WF defenses, and we expect our results may stir follow-up research into its robustness against such safeguards (e.g., by triggering potential extensions to its model architecture).

\mypara{Overlapped traces.} Lastly, in Table~\ref{tab:overlaps}, we compare the accuracy obtained by TSA-WF (w/ STUMPY) and ARES when identifying a monitored website amongst 3, 5, and 7 tab traces that have been overlapped to different degrees. For a (maximum) 10\% overlap, the accuracy of TSA-WF decreased (from 50.5\% and 41.2\%) by 5.0\% in the \texttt{3-Tab} setting and by 6.8\% in the \texttt{5-Tab} setting. In ARES, we observed a slightly larger decrease of 6.9\% in the \texttt{3-Tab} setting and 9.1\% in the \texttt{5-Tab} setting. As expected, we can see a trend that larger overlaps between traces lead to an overall decrease in classification accuracy.

\begin{figure}[t!]
    \centering
    \begin{minipage}{0.6\textwidth}
        \centering
        \raggedright
        \includegraphics[width=0.95\linewidth]{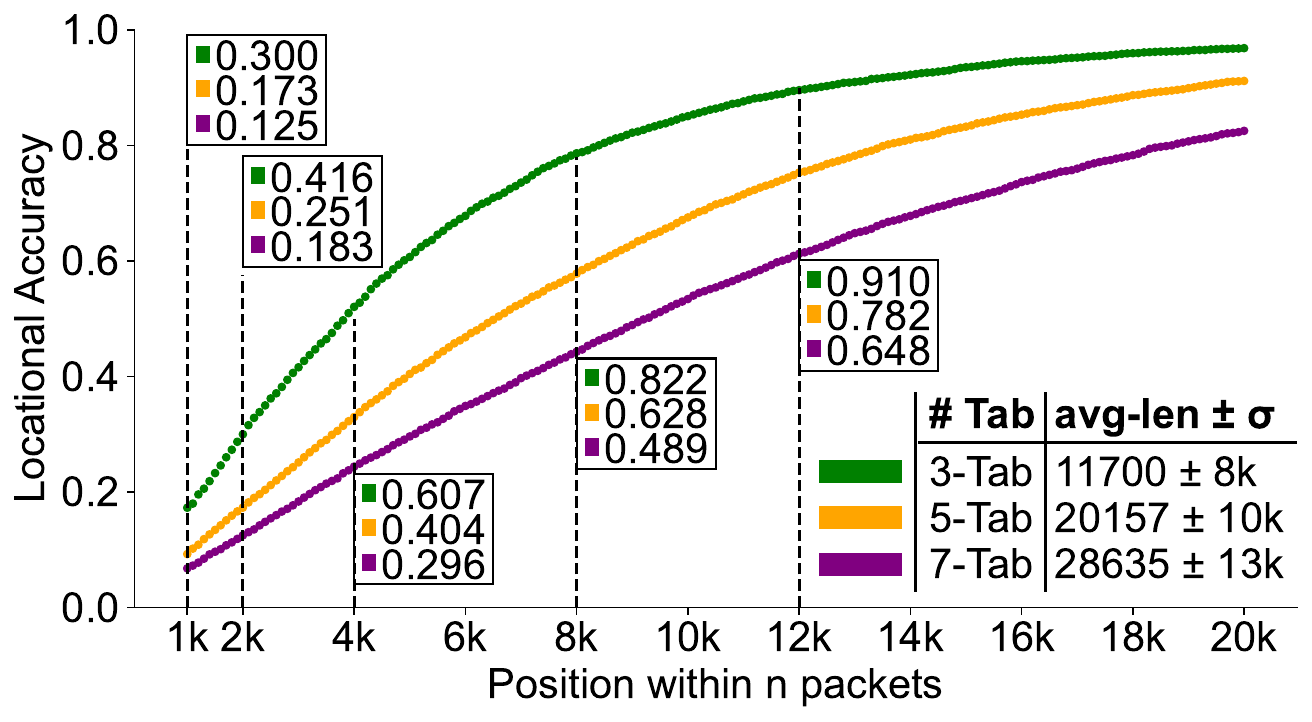}
        \textbf{Figure 8:} TSA-WF's accuracy at locating the monitored website within $n$ packets of a trace.
    \end{minipage}%
    \hfill
\begin{minipage}{0.38\textwidth}
    {\textbf{Table 8:} Execution time for TSA-WF and other WF attacks.}
    \par\addvspace{0.2cm}
    \scriptsize
    \begin{tabular}{@{}lcccc@{}}
        \toprule
        \textbf{Attack} & \multicolumn{4}{c}{\textbf{\# of Tabs}} \\ \cmidrule(l){2-5}
        & \textbf{1-Tab} & \textbf{3-Tab} & \textbf{5-Tab} & \textbf{7-Tab} \\ \midrule
        TSA-WF & 1 & \textbf{4} & \textbf{7} & \textbf{10} \\
        Euclidean & 10 & 35 & 44 & 60 \\
        WED & 13 & 36 & 47 & 62 \\
        CBD & 32 & 86 & 120 & 192 \\ \midrule 
        ARES & - & 10 & 10 & - \\
        Tik-Tok & 1.3 & - & - & - \\ 
        DF & 1.3 & - & - & - \\
        k-FP & \textbf{0.7} & - & - & - \\ \bottomrule
    \end{tabular}
\end{minipage}
\end{figure}

\subsection{Untangling Multi-tab Traces}
\label{sec:improve-dl}

In Figure 8, we depict the accuracy of TSA-WF when pinpointing the approximate location of a monitored website within an open-world multi-tab trace. For each unlabeled sample, we begin by taking the prototype associated with the monitored class label predicted by TSA-WF.  Then, we infer that monitored website's location in the multi-tab trace by finding the packet index where the minimum distance occurs (i.e., best match) between itself and the multi-tab trace being labeled.  The reported accuracy value represents the percentage of monitored websites which were correctly located to within $n$ packets of the true location.  To further understand these results, we note that the average length of \texttt{1-Tab} traces is $\bar{x}=3800$ with a standard deviation of $\sigma=4275$.  The 95th percentile length was 12145 packets, and the single largest trace in the dataset had 50728 packets.  In the bottom right of Figure 8, we also report the average length of multi-tab traces, 11700 for \texttt{3-Tab}, 20157 for \texttt{5-Tab}, and 28635 for \texttt{7-Tab}.

Overall, TSA-WF was able to correctly identify the location of the monitored website with an accuracy between 30.0\% and 91.0\% for \texttt{3-Tab} traces, 17.3\% and 78.2\% for \texttt{5-Tab} traces, and 12.5\% and 64.8\% for \texttt{7-Tab} traces. In the \texttt{3-Tab} setting, TSA-WF correctly located the monitored website to within 4\,000 packets with an accuracy of 60.7\%, which is similar to its overall accuracy on undefended \texttt{3-Tab} traces (52.2\%) with a difference of $+$8.5\%.  In the \texttt{5-Tab} and \texttt{7-Tab} settings, TSA-WF located the monitored website to within 4\,000 packets with an accuracy of 40.4\% ($-$0.8\% vs. undefended) and 29.6\% ($-$5.5\% vs. undefended).  We posit that as the number of tabs increases, the ability for TSA-WF to locate the website of interest to within one trace's length decreases faster than its classification accuracy, likely due to the increased chance for mismatches as a larger number of unmonitored websites are added to the merged trace.

In summary, TSA-WF can find the approximate location of a monitored website in a multi-tab trace, enabling adversaries to: a) pass the subsequence of the unlabeled sample which contains a monitored trace into single-tab WF attacks for better accuracy; b) feed the subsequence back into TSA-WF (or other WF attack) as an additional prototype or as supplementary training data.

\subsection{Computational Efficiency of TSA-WF}
\label{sec:compefficiency}

In Table 8, we first compare the total runtime of four different time series analysis techniques compatible with TSA-WF. Overall, STUMPY outperformed all other techniques by a factor of at least 5-6x.  The majority of our experiments throughout \S\ref{sec:experiment} (excluding Table \ref{tab:main-result}) were performed using only STUMPY, further demonstrating that its performance optimizations do not compromise accuracy.

Table 8 also compares the total runtime of TSA-WF against existing WF attacks, and we found it to be slower than k-FP but comparable to DF and Tik-Tok in the \texttt{1-Tab} setting. Furthermore, in the multi-tab setting, TSA-WF outperforms ARES' total runtime in both the \texttt{3-Tab} and \texttt{5-Tab} setting (4h and 7h for TSA-WF, respectively vs. 10h for ARES). While completing a full execution round faster than ARES, we note that the complete testing time for TSA-WF (i.e., prediction of 5\,000 samples, given our dataset as described in \S\ref{sec:goals-and-method}) is roughly 1/9th of the reported execution time from Table 8 (i.e., 26min for \texttt{3-Tab} and 46min for \texttt{5-Tab}), as time series distances need to be computed for each sample under prediction. In contrast,  ARES' testing time is approximately 9s in both the \texttt{3-Tab} and \texttt{5-Tab} setting---note this value is similar for both settings since ARES operates on fixed-sized inputs (see \S\ref{sec:multi-tab-experiment}).

\section{Conclusions}
In this paper, we characterized encrypted website traces as time series data, and reframed WF attacks as a time series matching problem.
This enabled the development of TSA-WF, a time series analysis pipeline specifically designed for WF. Our evaluation showed that TSA-WF achieves similar performance to state-of-the-art attacks in the open-world single-tab setting, both for undefended and defended website traces. Furthermore, we described how TSA-WF can augment the capabilities of existing multi-tab WF attacks by pinpointing the approximate instant at which a monitored website is accessed within a multi-tab trace. 

\section{Future work} Time series analysis remains a thriving research area. Given TSA-WF's flexibility with time series distance metrics, exploring recent matching techniques~\cite{middlehurst2024bake} could further enhance it.Another facet of TSA-WF that could be expanded upon is its potential ability to perform a stream-based analysis of network traces. Similarly to Holmes~\cite{10.1145/3658644.3670272}, TSA-WF offers the possibility to perform early-stage trace evaluation using the sliding window approach described in \S\ref{sec:distance}. This could forgo extensive pre-training of complex WF attack models, e.g., by using a simple per-class time series distance threshold (i.e., eschewing the training of the ML classifier currently in use by TSA-WF's prediction phase). 

\raggedbottom
\pagebreak

\bibliographystyle{splncs04}  
\bibliography{short-refs}

\end{document}